\documentclass[12pt]{article}
\usepackage{times}
\usepackage{graphicx}
\usepackage{geometry}
\usepackage[utf8]{inputenc}
\usepackage{enumitem,amssymb}
\usepackage{ragged2e}
\newlist{thematic}{itemize}{8}
\setlist[thematic]{label=$\square$}
\usepackage{pifont}
\newcommand{\cmark}{\ding{51}}%
\newcommand{\done}{\rlap{$\square$}{\raisebox{2pt}{\large\hspace{1pt}\cmark}}%
\hspace{-2.5pt}}

\usepackage{jheppub}   
\usepackage[dvipsnames,svgnames,x11names]{xcolor}
\usepackage{sectsty}
\usepackage[outercaption]{sidecap}
\sidecaptionvpos{figure}{c}

\newcommand{\arcmin}{\hbox{$^\prime$}}               % arcmin
\definecolor{DarkGreen}{rgb}{0.0, 0.3, 0.0}
\definecolor{purple}{rgb}{0.5, 0.0, 0.5}
\definecolor{red}{rgb}{1, 0.0, 0.0}
\definecolor{green}{rgb}{0, 1.0, 0.0}

\def\3he{$^3{\rm He}$}

\def\lsim{\mathrel{\lower2.5pt\vbox{\lineskip=0pt\baselineskip=0pt
           \hbox{$<$}\hbox{$\sim$}}}}

\def\gsim{\mathrel{\lower2.5pt\vbox{\lineskip=0pt\baselineskip=0pt
           \hbox{$>$}\hbox{$\sim$}}}}

\begin{document}
\raggedright
\huge
Astro2020 Science White Paper \smallskip

The case for a `sub-millimeter SDSS': a 3D map of galaxy evolution to $z\approx10$ 

%\linebreak

\bigskip

\normalsize

\noindent \textbf{Thematic Areas:} \hspace*{60pt} $\square$ Planetary Systems \hspace*{10pt} $\square$ Star and Planet Formation \hspace*{20pt}\linebreak
$\square$ Formation and Evolution of Compact Objects \hspace*{31pt} $\done$
 Cosmology and Fundamental Physics \linebreak
  $\square$  Stars and Stellar Evolution \hspace*{1pt} $\square$ Resolved Stellar Populations and their Environments \hspace*{40pt} \linebreak
$\done$
    Galaxy Evolution   \hspace*{45pt} $\square$             Multi-Messenger Astronomy and Astrophysics \hspace*{65pt} \linebreak
  
\textbf{Principal Author:}

Name:\,James~E.~Geach
 \linebreak						
Institution:\,Centre for Astrophysics Research, University of Hertfordshire, UK
 \linebreak
Email:\,j.geach@herts.ac.uk
 \linebreak
Phone:\,+44(0)1707\,281\,065

\vspace{-4pt}
\justify

\textbf{Co-authors:~}Manda~Banerji (Cambridge), 
Frank~Bertoldi (AIfA),                   
Matthieu~B\'ethermin (LAM),
Caitlin~M.~Casey (UT Austin), 
Chian-Chou~Chen (ESO),
David~L.~Clements (Imperial), 
Claudia~Cicone (INAF),  
Francoise~Combes (Obs-Paris), 
Christopher~Conselice (Nottingham),
Asantha~Cooray (UC Irvine),
Kristen~Coppin (Herts),
Emanuele~Daddi (CEA),
Helmut~Dannerbauer (IAC, ULL),
Romeel~Dav\'e (Edinburgh), 
Matthew~Doherty (Herts),
James~S.~Dunlop (Edinburgh),
Alastair~Edge (Durham),
Duncan~Farrah (Hawaii), 
Maximilien~Franco (CEA),
Gary~Fuller (Manchester), 
Tracy~Garratt (Herts),
Walter~Gear (Cardiff), 
Thomas~R.~Greve (UCL, DAWN),
Evanthia~Hatziminaoglou (ESO),
Christopher~C.~Hayward (Flatiron Institute),
Rob~J.~Ivison (ESO),
Ryohei~Kawabe (NAOJ),
Pamela~Klaassen (STFC), 
Kirsten~K.~Knudsen (Chalmers),
Kotaro~Kohno (U. Tokyo),
Maciej~Koprowski (UMK),
Claudia~D.~P.~Lagos (ICRAR),
Georgios~E.~Magdis (NBI), 
Benjamin~Magnelli (AIfA),
Sean~L.~McGee (Birmingham),
Micha{\l}~Micha{\l}owski (AMU, Pozna{\'n}), 
Tony~Mroczkowski (ESO),
Desika~Narayanan (Florida),
Omid Noroozian (NRAO, NASA),
Seb Oliver (Sussex), 
Dominik~Riechers (Cornell, MPIA),
Wiphu~Rujopakarn (Chulalongkorn), 
Douglas~Scott (UBC), 
Stephen~Serjeant (Open U.),
Matthew~W.~L.~Smith (Cardiff),
Mark~Swinbank (Durham),
Yoichi~Tamura (Nagoya),
Paul~van~der~Werf (Leiden),
Eelco~van~Kampen (ESO),
Aprajita~Verma (Oxford),
Joaquin~Vieira (Illinois),
Jeff~Wagg (SKA),
Fabian~Walter (MPIA, NRAO),
Lingyu~Wang (SRON), 
Al~Wootten (NRAO),
Min~S.~Yun (UMass)

\vspace{6pt}

\noindent \textbf{Abstract:~}The Sloan Digital Sky Survey (SDSS) was revolutionary because of the extraordinary breadth and ambition of its optical imaging and spectroscopy. We argue that a `sub-millimeter SDSS' -- a sensitive large-area imaging+spectroscopic survey in the sub-mm window -- will revolutionize our understanding of galaxy evolution in the early Universe. By detecting the thermal dust continuum emission and atomic and molecular line emission of galaxies out to $z\approx10$ it will be possible to measure the redshifts, star formation rates, dust and gas content of hundreds of thousands of high-{\it z} galaxies down to $\sim$$L_\star$. Many of these galaxies will have counterparts visible in the deep {\it optical} imaging of the Large Synoptic Survey Telescope. This 3D map of galaxy evolution will span the peak epoch of galaxy formation all the way back to cosmic dawn, measuring the co-evolution of the star formation rate density and molecular gas content of galaxies, tracking the production of metals and charting the growth of large-scale structure.

\pagebreak

\section{Introduction}

Sub-millimeter (sub-mm) surveys offer a unique and efficient probe of galaxy evolution \cite{SIB97,Hughes98}. Half of the radiation emitted by stars over all cosmic time has been absorbed by interstellar dust \cite{Dole06,Hill18}, which thermally re-radiates this energy approximately as a blackbody of temperature 30--50\,K, peaking in the rest-frame far-infrared (FIR) \cite{CNC14}. Cosmic expansion redshifts the bulk of the thermal dust emission into the sub-mm/mm window. This leads to the remarkably useful coincidence that, for a source of fixed IR luminosity, cosmological dimming of the observed flux density is effectively cancelled out as the rest-frame frequency probed by a sub-mm bandpass `climbs' up the Rayleigh-Jeans tail of thermal dust emission at increasing redshift \cite{Blain02} (Fig~1). This negative {\it K}-correction allows sub-mm surveys to detect galaxies out to $z\sim10$ with a nearly uniform luminosity limit for a given survey sensitivity, probing a huge cosmic volume. However, current and near-future  sub-mm/mm surveys suffer from at least one of the following shortcomings.

\smallskip

\noindent {\bf Poor resolution:}~ground-based single-dish sub-mm facilities have been limited to apertures of 15\,m, corresponding to beams typically exceeding $10''$, an order of magnitude larger than uncorrected seeing in the UV/Optical/IR (UVOIR) bands. This has two negative impacts:~(i) correct identification of the corresponding source, or sources, in other bands can be challenging due to multiple potential counterparts \cite{Ivison97,An19}; and (ii) confusion noise \cite{Helou90} due to the presence of multiple unresolved sources crowding the large beam. Confusion implies a sensitivity floor, beyond which the noise no longer scales with $\smash{t^{-1/2}}$. For current surveys, this typically results in samples of galaxies in the ultraluminous ($L_{\rm IR}\gtrsim10^{12}L_\odot$) class, with star formation rates (SFRs) of 100s\,$M_\odot$\,yr$^{-1}$. {\it Solution:~build a large aperture single-dish facility to reduce confusion noise in the sub-mm.}
   
\smallskip
   
\noindent {\bf Small survey areas:}~although sub-mm detector technology has advanced in recent years, allowing for the manufacture of large-format bolometer arrays (e.g.,\ \citep{Holland13}), mapping efficiency is still limited. This has restricted the scope of deep (i.e.,\ $S_\nu\sim1$\,mJy depth) extragalactic surveys conducted from the ground in the sub-mm to $\sim$10\,deg$^2$ \cite{Weiss09,Geach13, Geach17,Wang17}, i.e.,\,orders of magnitude smaller than modern and near-future UVOIR and radio surveys that cover 1000s of square degrees \cite{DES}. {\it Herschel} has mapped large areas at 250/350/500$\mu$m \cite{HATLAS,HERMES}, but its 3.5-m mirror resulted in very large beams and therefore high confusion. Similarly, ground-based cosmological surveys in the sub-mm/mm, such as the South Pole Telescope (SPT, \cite{SPT}) and Atacama Cosmology Telescope (ACT, \cite{ACT}), {\it have} mapped 1000s of square degrees, but with a focus on cosmic microwave background (CMB) science, and are not deep enough to detect any but the brightest galaxies (e.g.,\ blazars and strongly lensed galaxies \cite{Vieira13}). {\it Solution:~develop large format cameras with a mapping efficiency capable of surveying $\sim$1000\,deg$^2$ to low confusion limits in a reasonable ($\sim$1000\,hour) time.}
 
\smallskip

\noindent  {\bf Lack of spectroscopic component:}~The UVOIR counterparts to sub-mm galaxies are often faint or undetected, or lie at redshifts with no easily accessible emission lines in the UVOIR windows (e.g.,\ the redshift desert, \cite{Steidel04}). Combined with the points above, there is currently no efficient coupling of highly multiplexed spectroscopic follow-up to sub-mm galaxy surveys. Although the FIR/sub-mm/mm is rich in spectral features, arising from the chemistry of the cool interstellar medium (e.g.,\ the bright carbon monoxide (CO) rotational ladder and the FIR atomic fine-structure lines ([C~{\sc ii}], [N~{\sc ii}]), current spectrometers operating at these frequencies are: (a) not multiplexed and (b) either lack the sensitivity or wide bandwidth for efficient spectroscopic identification of the 1000s of sources detected in continuum surveys. {\it Solution:~develop on-chip spectrometers to turn the the large format cameras described above into broadband integral field units.}

\vspace{-2mm}
\begin{minipage}[h]{0.45\textwidth}\vspace{0.2cm}
\hspace{-1.3cm}\includegraphics[width=\textwidth]{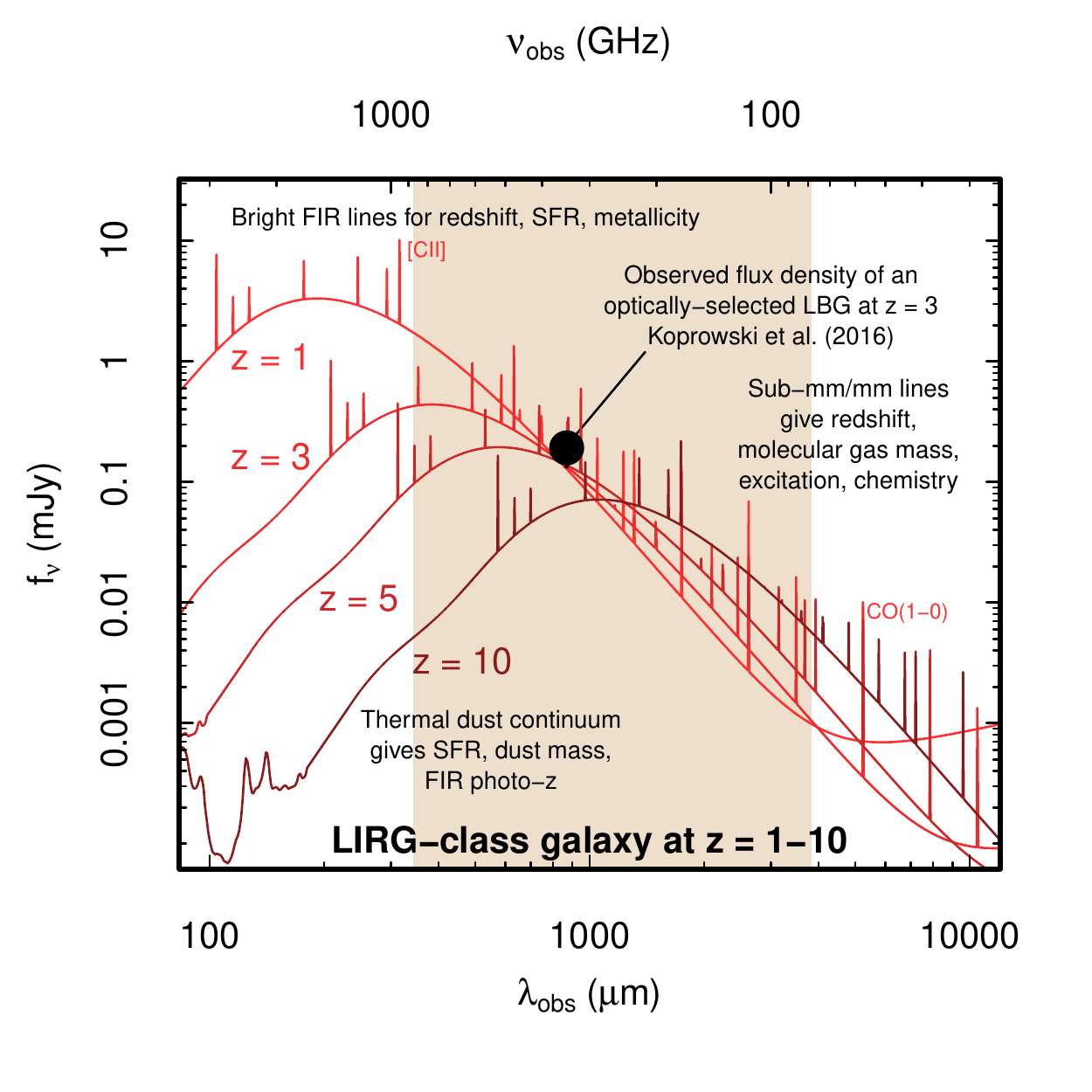}
\end{minipage}\hspace{-1cm}\begin{minipage}[h]{0.575\textwidth}
{\bf Fig.~1~}The spectral energy distribution of a star-forming galaxy at $z=1$--$10$, scaled to $L_{\rm IR}=10^{11}L_\odot$. The template is based on the well-studied `Cosmic Eyelash' \cite{Eyelash}. The point shows the directly detected 870$\mu$m continuum emission of an {\it optically-selected} Lyman-break galaxy at $z=3$ with a UV+IR SFR of 35\,$M_\odot$\,yr$^{-1}$ \cite{Koprowski16}. The rich spectral landscape in the sub-mm/mm opens up the possibility of measuring spectroscopic redshifts for hundreds of thousands of high-{\it z} galaxies that will {\it also} be detected by LSST. 
\end{minipage}
\vspace{-8mm}

\section{A Sub-millimeter Spectroscopic Survey}

The Sloan Digital Sky Survey (SDSS \citep{York00}) has demonstrated the power of a large imaging survey coupled with a dedicated spectroscopic component; over nearly two decades it has enabled a remarkably diverse range of science and has a continuing legacy with on-going projects. Here we argue that a `sub-mm SDSS' using a new 50-m-class single-dish sub-mm telescope would be a similarly powerful and unique tool for galaxy evolution studies. Accessing the sub-mm sky is essential for two reasons: (i) probing rest-frame frequencies close to the peak of the thermal dust emission to obtain accurate estimates of the total $L_{\rm IR}$ and dust temperature; and (ii) targeting the key diagnostic FIR lines \cite{Stacey93} across all redshifts, from the peak epoch of galaxy formation at $z\approx1$--$2$ to earlier times (e.g.,\ [C~{\sc ii}] is observed at 474\,$\mu$m at $z=2$). We envision a simple, but highly versatile large-area blind imaging and spectroscopic survey with strong legacy value in and of itself and with a high degree of complementarity to existing and future projects such as Large Synoptic Survey Telescope (LSST, \citep{lsst}), {\it Euclid}, \cite{euclid}, {\it WFIRST} \cite{wfirst}, {\it SPHEREx} \cite{spherex}, {\it Athena} \cite{athena} and Square Kilometer Array (SKA) surveys.  To lay out the broad science themes,  feasibility and basic requirements of such a survey, let us focus on a new facility called the Atacama Large-Aperture Sub-mm/mm Telescope (AtLAST)\footnote{\url{http://atlast-telescope.org}}. The ideas presented in this White Paper find their roots in the science drivers behind previous (and ongoing) efforts to design and build a new large single-dish sub-mm facility such as CCAT \cite{CCAT} and the Large Sub-millimeter Telescope (LST \citep{Kawabe17}), and emphasizes a continuing international appetite for such a telescope. 

\subsection{Key science}

The 2-Gyr window spanning $z\approx1$--$10$ encompasses the `cosmic dawn' when galaxies first formed and reionized the Universe, to `cosmic noon', as marked by the apex of the cosmic star formation rate density (SFRD, \citep{MD14}). Although the basic picture of a rapid ramp-up of the SFRD from $z\approx10$ is now roughly constrained from an observational standpoint, these insights are almost exclusively biased towards deep UVOIR imaging surveys that estimate the UV luminosity function by identifying Lyman drop-out systems at high-{\it z} \cite{Bouwens15}. In contrast, our knowledge of the dust-obscured early Universe, and the evolution of the cold and dense interstellar medium (ISM) is lacking. A comprehensive sub-mm spectroscopic survey will transform our understanding of the early phases of galaxy formation, performing several key observations that will allow dramatic breakthroughs:

\vspace{-0.5cm}
\begin{center}
\begin{tabular}{|p{0.475\textwidth}|p{0.475\textwidth}|}
    \hline  
    {\bf Observation} & {\bf Breakthrough} \cr
    \hline
    Infrared luminosities and SFRs of galaxies (dust continuum, lines) & Perform a complete census of star-forming galaxies at high-{\it z} to sub-$L_\star$ luminosities \cr
    \hline
    Dust content of galaxies (dust continuum) & Reveal the production and evolution of metals in the Universe, as tracked by the dusty ISM\cr
    \hline
    Molecular gas content of galaxies (CO lines) & Determine the evolution of the co-moving H$_2$ mass density and the astrophysics governing star formation efficiency and ISM chemistry \cr
    \hline
    Clustering of star-forming galaxies (angular and redshift-space correlation functions) & Chart the growth of large scale structure at the same time galaxies are being assembled; detect baryonic acoustic oscillations beyond $z\gtrsim2$  \cr
    \hline
\end{tabular}
\end{center}

\noindent We keep in mind that many of the driving questions pertinent to the field of galaxy formation a decade hence have yet to be asked. This demands a versatile survey with broad scope, to be exploited by future astronomers who will answer questions {\it not yet} articulated by the community.

\subsection{The State Of The Art}
 
 In the sub-mm, the SCUBA-2 camera on the 15-m James Clerk Maxwell Telescope (JCMT) has been one of the workhorse instruments for large extragalactic imaging surveys in the sub-mm \citep{Holland13}. At longer wavelengths\footnote{Discounting the very large cosmological surveys of SPT and ACT.}, the New IRAM KID Array (NIKA \citep{nika}, now upgraded to NIKA-2) camera on the 30-m IRAM telescope offers imaging at 1.25 and 2.14 mm using an array of $\sim$100\,mK cooled kinetic inductance detectors (KIDs) \cite{Day2003}. KID arrays are becoming established as the detector of choice for large-format sub-mm/mm imaging \cite{Sayers2016,Baselmans2017A&A,Austermann2018}. For example, the 50-meter Large Millimeter Telescope (LMT) has recently started science operations. TolTEC, the new imaging polarimeter developed for LMT, employs arrays of 1000s of KIDs, with projected mapping speeds of over 2/3/10\,deg$^2$\,mJy$^2$\,hr$^{-1}$ at 1.1/1.4/2\,mm \cite{Bryan2018}. With a beam under 10$''$ for $\lambda<2$\,mm, LMT's large aperture results in a sub-mJy confusion limit and therefore probes a more representative sample of mm-selected galaxies. However, the LMT site (Sierra Negra, 4850\,m) and primary surface are suboptimal for frequencies $\nu_{\rm obs}>350$\,GHz, and LMT's field of view (FoV) is limited to 4$\arcmin$.

KIDs also offer a promising route to constructing large-format imaging spectrometers. For example, the Deep Spectroscopic High-redshift Mapper (DESHIMA \cite{Endo2018,Endo2012}) is a new broadband spectrometer operating at 346-GHz with an instantaneous bandwidth of 40\,GHz, recently tested on the Japanese Atacama sub-millimeter Telescope Experiment (ASTE). MOSAIC will be a successor instrument, scaling DESHIMA's single pixel to a 25-pixel array, with potentially an even wider bandwidth to cover 240--720\,GHz. Other `on-chip' sub-mm spectroscopic solutions with moderate spectral resolution include WSPEC \cite{Bryan2016}, $\mu$Spec \cite{Noroozian2015, Barrentine2016,Cataldo2018}, SuperSpec \cite{Hailey-Dunsheath2016}, and the CAMbridge Emission Line Surveyor (CAMELS) \cite{Thomas2014}. Thus, wideband imaging-spectrometers will soon become a reality in the sub-mm. One could envision a large format KID-based imaging spectrometer that could map large areas to simultaneously measure continuum and line emission -- and therefore immediately measure the redshifts -- of 1000s of high-{\it z} galaxies. With spectral information, one can overcome some of the confusion issues that hamper single-band continuum imaging, such that for a given aperture the effective confusion limit of a spectral/multi-band survey is lower than a single-band continuum survey. Still, it is desirable to achieve as low a confusion limit as possible in order to probe `normal' ($L_\star$) galaxies at all redshifts. Finally we note that the 6-m CCAT-prime telescope \cite{CCATp} (located at 5600\,m, above the ALMA array) is expected to be operational in 2021. With its 8$^\circ$ FoV and 350\,$\mu$m--2\,mm spectral coverage, CCAT-prime will conduct spectroscopic and photometric surveys with a spatial resolution significantly better than {\it Herschel}, but still not detecting individual $L_\star$ galaxies at high-{\it z}.

\subsection{Opportunities to advance the field}

The science goals described above could be achieved with a new survey facility that would combine a 50-m-class single-dish telescope with a large-format KID-based imaging spectrometer, comprising 1000s of pixels over an instantaneous FoV of 1$^\circ$ and covering a bandwidth of 80--720\,GHz. The combination of large FoV and broad spectral coverage is key for efficiently mapping large areas of sky and providing access to a wide range of redshifted spectral features from galaxies spanning a huge cosmological volume. To operate efficiently at high-frequencies, this facility would need a high altitude site, with the 5100\,m Chajnantor plateau being the obvious choice, not just considering the clear synergies with ALMA, but also taking into account the existing infrastructure established on the plateau. Other sites might be possible: pushing to higher altitude (like CCAT-prime) would be better for $\lambda_{\rm obs}\leq450$\,$\mu$m observing (and potentially allow for a smaller dish if there were significant gains in efficiency) compared to the plateau, but with significant challenges in construction, operation and overall cost. ALMA itself could never achieve a blind imaging--spectroscopic survey spanning 1000\,deg$^2$; indeed, currently it remains observationally inefficient just to determine the redshift of a single continuum-detected source. ALMA's strengths lie in detailed and sensitive follow-up studies, so pairing with a dedicated survey (such as the one we propose) would provide a significant boost to the efficiency of ALMA's exploration of the early Universe.   

A 50-m single-dish delivers a 1.8$''$ beam at 450$\mu$m, allowing one to pin-point a 5$\sigma$ continuum point-source detection to within 0.2$''$ -- entirely adequate for accurately associating sub-mm sources with their UVOIR counterparts. A major breakthrough would be the routine detection and redshift identification of high-{\it z} star-forming galaxies that will also be detected in deep UVOIR imaging surveys, {\it as well as} the highly obscured and therefore optically-dark galaxies that will be missed in such surveys. For example,  optically-selected ($i<27$\,mag) Lyman-break galaxies (LBGs) at $z\approx3$--$5$ have been measured to be in the LIRG-class ($L_{\rm IR}\sim10^{11}L_\odot$) and have 850-$\mu$m flux densities of order 100\,$\mu$Jy \cite{Coppin15,Koprowski16} (Fig~1). Therefore, an instrument that could achieve broadband (e.g.,~8\,GHz aggregate) sensitivities of order 20\,$\mu$Jy at 850-$\mu$m could directly detect the thermal dust emission from the {\it same} galaxies from which the UV luminosity function is derived (at least down to $\sim$$L_\star$) in the key `ramp-up' epoch of galaxy growth, thus performing a complete census of star formation at early times.

Multiple continuum measurements across the sub-mm/mm window alone could provide far-infrared photometric redshift estimates (which could be combined with UVOIR photometry), however with spectroscopic capability one could also directly detect multiple atomic and molecular emission lines to measure precise redshifts. Consider one of the brightest fine-structure lines, [C~{\sc ii}]$\lambda158~\mu$m: this line typically carries between approximately 0.1--1\% of $L_{\rm IR}$ \cite{Malhotra01,Luhman03}. Assuming a conservative $\log(L_{\rm [CII]}/L_{\rm IR})=-2.5$ \cite{Stacey10}, the expected line flux averaged over a 200\,km\,s$^{-1}$ line (again, probably a conservative assumption, given the [C~{\sc ii}] line widths observed in some high-{\it z} galaxies, e.g.,~\cite{Smit18}) is approximately 5(2)\,mJy for a LIRG at $z=3(5)$. Our spectroscopic survey would trade spectral resolution for sensitivity, such that we would not be aiming to resolve emission lines; direct follow-up with ALMA -- having established the redshift and total line flux -- would be a powerful synergy here. Therefore, the instrument would need to achieve a 1$\sigma$ sensitivity of order 500\,$\mu$Jy per 200\,km\,s$^{-1}$ channel in reasonable integration times to efficiently detect emission lines in galaxies with SFRs of order 10s\,$M_\odot$\,yr$^{-1}$ at these redshifts.

We predict the number counts of CO and [C~{\sc ii}] line emitters detected in a blind spectroscopic survey in the sub-mm/mm windows by considering a conservative model for the minimal emergent luminosity of each line that is dependent on $L_{\rm IR}$ and assuming a model for the evolving infrared luminosity function \cite{Bethermin17}. We estimate that a deep blind spectroscopic survey (down to the confusion of a 50-m antenna, approximately 100\,$\mu$Jy at 1\,mm) will provide multiple line detections for about 150,000 galaxies per square degree (see also \citep{GP12}) out to at least $z\approx7$ (although note that for increasing redshift one also needs to take into account the contrast of sub-mm/mm lines against the CMB \citep{dC13}). Such a genuine high-{\it z} survey capability is far beyond the reach of ALMA and any planned UVOIR facilities. 

\section{Conclusions}

A large-area, sensitive spectroscopic imaging survey in the sub-mm will provide a 3D map of galaxy evolution to $z\approx10$. The key technological breakthrough required will be the development of large arrays of on-chip spectrometers that can achieve sub-mJy sensitivity in channels of width $\sim$100\,km\,s$^{-1}$ across 80--720\,GHz. A 1000 pixel camera would require $\sim$10$^8$ detectors, such that the cost per detector would have to be brought below \$1 to make such an instrument financially viable. The arguments for a sub-mm spectroscopic survey will of course be familiar to astronomers already working in this waveband, but it is important to emphasize that such a survey will also be of profound relevance to the entire high-{\it z} community. For example, LSST will image the sky down to $i\lesssim28$\,mag, and therefore a large fraction of LSST-selected high-{\it z} galaxies could be directly detected in the thermal dust continuum {\it and} have secure spectroscopic redshifts from their FIR/sub-mm/mm lines. In conclusion, a `sub-millimeter SDSS' would have far-reaching cross-community relevance, opening up a vast discovery space, and directly delivering and helping to facilitate a diverse and rich range of science legacy for the coming decades.  

\pagebreak
%\textbf{References}

\bibliographystyle{unsrturltrunc6}
\bibliography{references}

\end{document}